\documentclass[twocolumn,showpacs,preprintnumbers,amsmath,amssymb]{revtex4}

\usepackage{graphicx}
\usepackage{dcolumn}
\usepackage{bm}

\newcommand{\eq}{\begin{equation}}
\newcommand{\eqx}{\end{equation}}
\newcommand{\eqn}{\begin{eqnarray}}
\newcommand{\eqnx}{\end{eqnarray}}

\begin{document}


\title{Ionization and expansion dynamics of atomic clusters 
irradiated with short intense VUV pulses}

\author{B. Ziaja \footnote{Corresponding author. E-mail: ziaja@mail.desy.de}, H. Wabnitz, F. Wang and E. Weckert}
\affiliation{Hamburger Synchrotronstrahlungslabor,\\
Deutsches Elektronen-Synchrotron, \\
Notkestr.\ 85, D-22603 Hamburg, Germany\\ 
}

\author{T. M\"oller}
\affiliation{Technische Universit\"at Berlin,\\
Institut f\"ur Atomare Physik und Fachdidaktik,\\
10623 Berlin, Hardenbergstrasse 36, Germany}

\date{\today}

\begin{abstract}
Kinetic Boltzmann equations are used to model the ionization and expansion dynamics of xenon clusters irradiated with short intense VUV pulses.
This unified model includes predominant interactions that contribute to the cluster dynamics induced by this radiation.
The dependence of the evolution dynamics on cluster size, $N_{atoms}=20-90000$, and pulse fluence, $F=0.05-1.5$ J/cm$^2$, corresponding to intensities
in the range, $10^{12}-10^{14}$ W/cm$^2$ and irradiation times, $\leq 50$ fs, is investigated. The predictions obtained with our model are found to be in good agreement with the experimental data. We find that during the exposure the cluster forms a shell structure consisting of a positively charged outer shell and a core of net charge equal to zero. The width of these shells depends on the cluster size.
The charged outer shell is large within small clusters ($N_{atoms}=20,70$), and its Coulomb explosion drives the expansion of these clusters. Within the large clusters ($N_{atoms}=2500,90000$) the
neutral core is large, and after the Coulomb explosion of the outer shell it expands hydrodynamically. Highly charged ions within the core recombine efficiently with electrons. As a result, we observe a large fraction of neutral
atoms created within the core, its magnitude depending on the cluster size.
\end{abstract}


\maketitle

\noindent

Atomic clusters are excellent objects to test the dynamics within samples irradiated with radiation from short wavelength free-electron-lasers (FELs) \cite{desy2006,slac,jap}. Their physical properties put them on the border between the solid state and the gas phase. Cluster studies are important for planned experiments with FELs in solid state physics, materials science and for studies of the extreme states of matter \cite{xfelinfo2007}.
Accurate predictions on the ionization, thermalization and expansion timescales within irradiated samples that can be obtained with cluster experiments are also needed for exploring the limits of experiments on single particle diffraction imaging \cite{l1,miao,gyula1,plasma4,liver1en,chapman}.

During the first cluster experiments performed at the free-electron-laser 
facility FLASH at DESY with VUV photons of energy, $E=12.7$ eV, and power densities up to a few $10^{13}$ W/cm$^2$ \cite{desy} highly charged Xe ions 
(up to $+8$) of high kinetic energies were detected. This unexpectedly strong energy absorption could not be explained using the standard approaches \cite{desy5,desy2,desy7}. More specifically, the energy absorbed was almost an order of magnitude larger than that predicted with classical absorption models, and the ion charge states created were much higher than those observed during the irradiation of isolated atoms in similar conditions. This indicated that at such radiation wavelengths some processes specific to many-body systems were responsible for an enhanced energy absorption. 

The physics underlying the dynamics within the irradiated clusters is complex. Several interesting theoretical models have been proposed in order to describe the evolution of clusters exposed to intense VUV pulses \cite{santra,santra1,siedschlag,georg,bauer,brabec,rusek}.
The most explored ones are: i) heating of quasi-free electrons due to enhanced inverse bremsstrahlung (IB) \cite{santra,santra1}, ii) enhanced photoionization within the sample due to lowering of interatomic potential barriers \cite{siedschlag,georg}, and iii) heating due to many-body recombination processes \cite{brabec}. Each of these various approaches lead to the
significantly enhanced energy absorption in agreement with the experimental data
\cite{desy}. However, one would expect that if all proposed enhancement factors are included within one model, it would probably lead to absorption rates much higher than those experimentally observed.

In order to evaluate the contribution of various processes to the ionization dynamics, we have constructed a unified model \cite{ziajab2}, \cite{ziajab3} based on kinetic equations including the following predominant interactions: photoionization, collisional ionization, elastic scattering of electrons on ions, inverse bremsstrahlung heating, electrostatic interactions between charges and with laser field, shifts of energy levels within atomic potentials due to the plasma environment, and shielded electron-electron interactions.
Using a non-equilibrium Boltzmann solver, we followed the evolution of the 
Xe clusters of various size ($N_{atoms}=20-90000$) irradiated with rectangular VUV pulses of intensity $10^{12}-10^{14}$ W/cm$^2$ and duration, $\leq 50$ fs.
We found that all physical mechanisms that were included into the model contributed to the ionization dynamics but with different weights. The total ionization rate within the sample was affected most by the inverse bremsstrahlung heating rate applied. Within the theoretical framework defined above we estimated that: i) many-body effects (many-body recombination) \cite{brabec} could contribute only for clusters irradiated at low pulse fluences, ii) the plasma environment effects estimated with electron screened atomic potentials were small. This was in contrast to the estimates of Ref. \cite{siedschlag} performed with unscreened atomic potentials. Our study extended the treatment of Ref.\ \cite{santra1}, where neither the spatial inhomogeneity of the clusters was treated nor the gross movement of electrons. It turned out that both effects significantly contribute to the cluster dynamics: the structure of charge created within the cluster was found to be strongly inhomogeneous \cite{desy8}. This inhomogeneity was induced by the dynamics of electrons. 

The main aim of this letter is to obtain a complete description of ionization and expansion dynamics. Therefore we have included into the present analysis
the evaluation of the effect of the three-body-recombination process (the non-equilibrium recombination rates are included into kinetic equations), and the expansion of the cluster until ions start to leave the simulation box (up to $\sim 2$ ps). 

\begin{figure*}
a) \includegraphics[width=5cm]{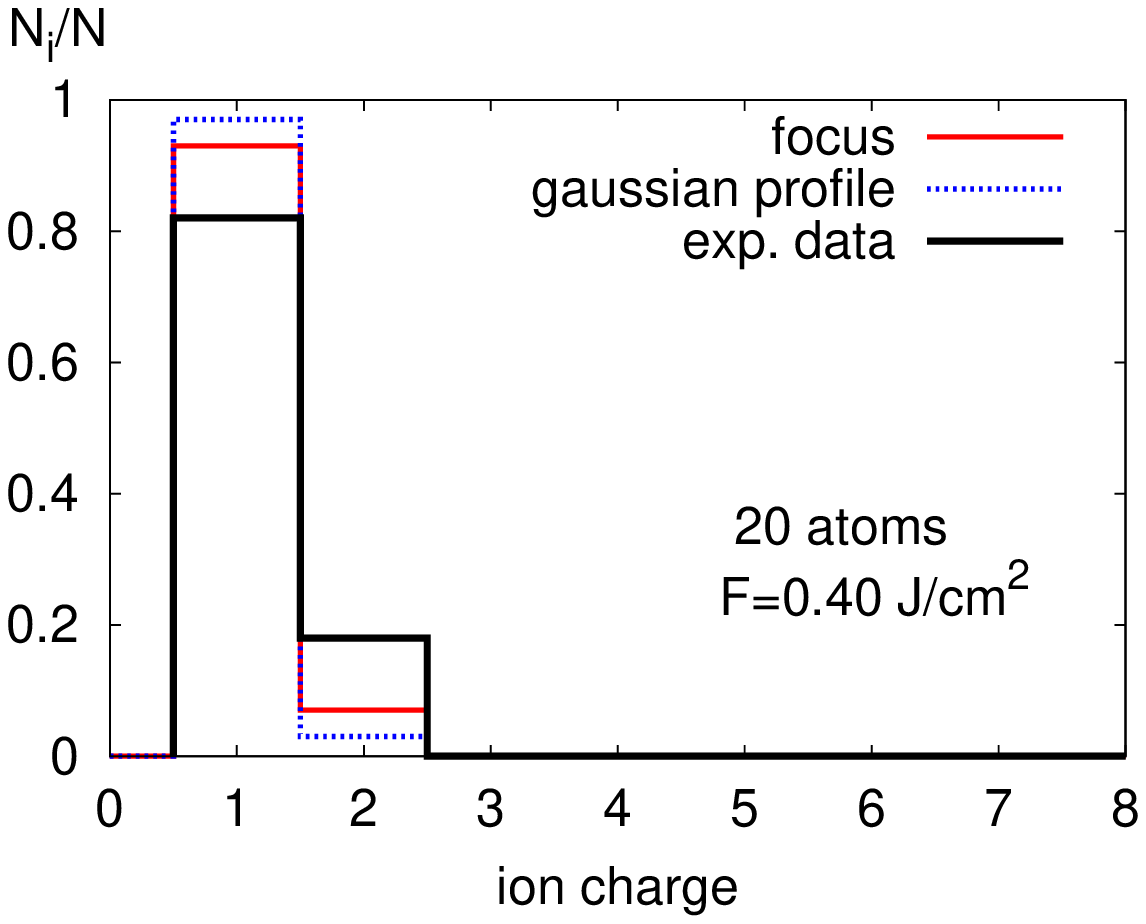}
c) \includegraphics[width=5cm]{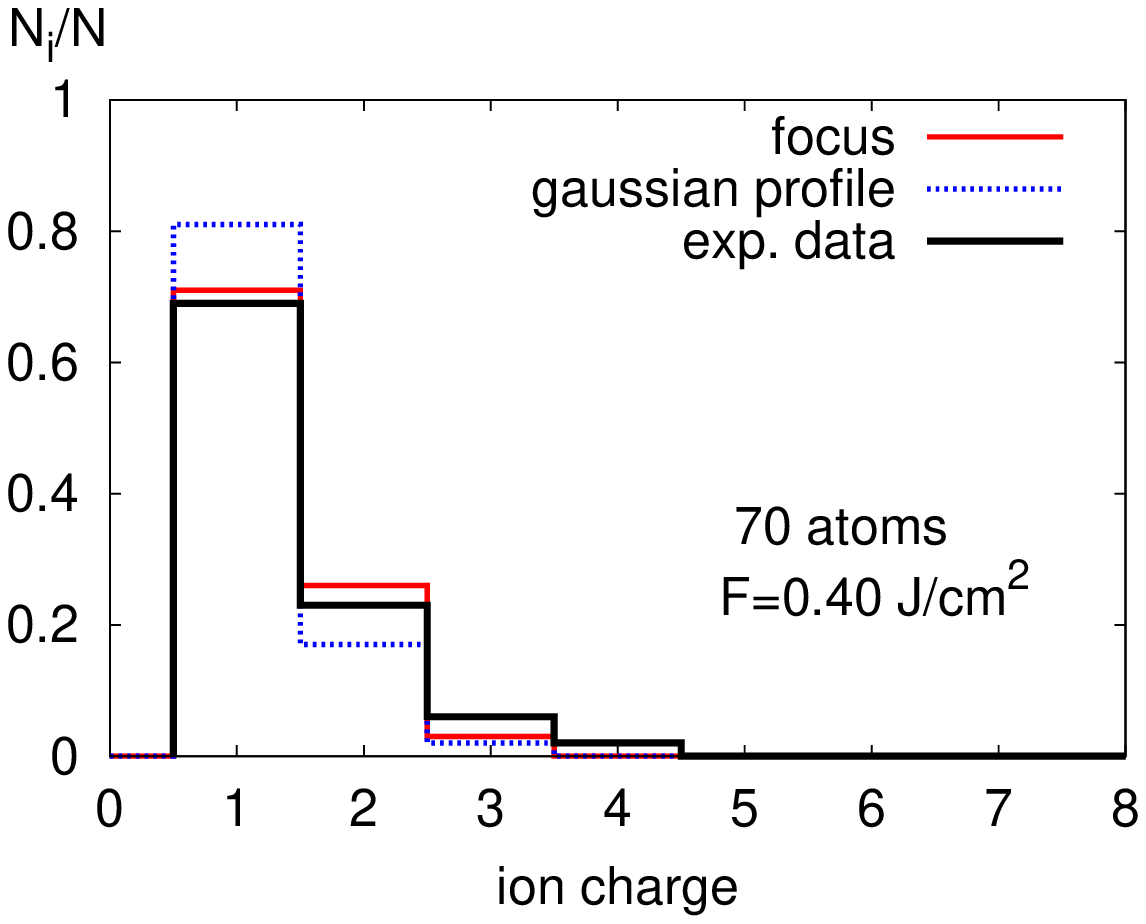}
e)\includegraphics[width=5cm]{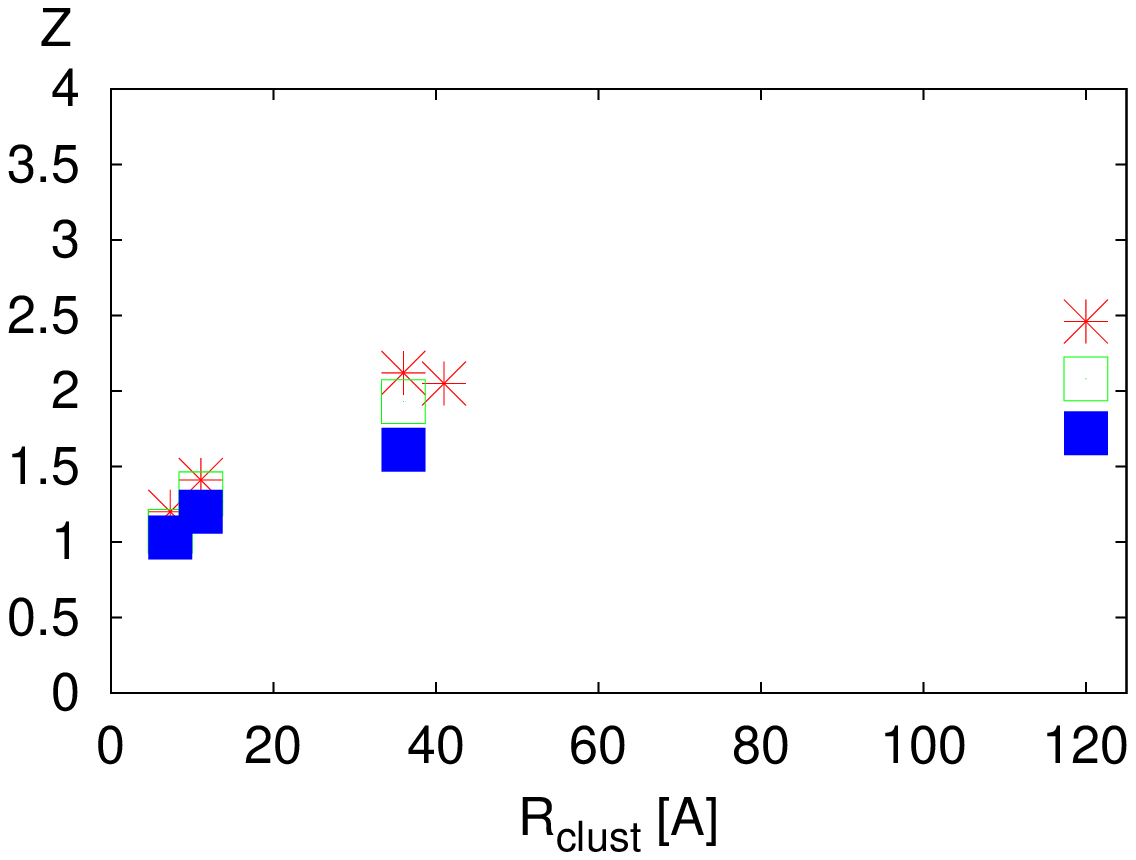}\\
b) \includegraphics[width=5cm]{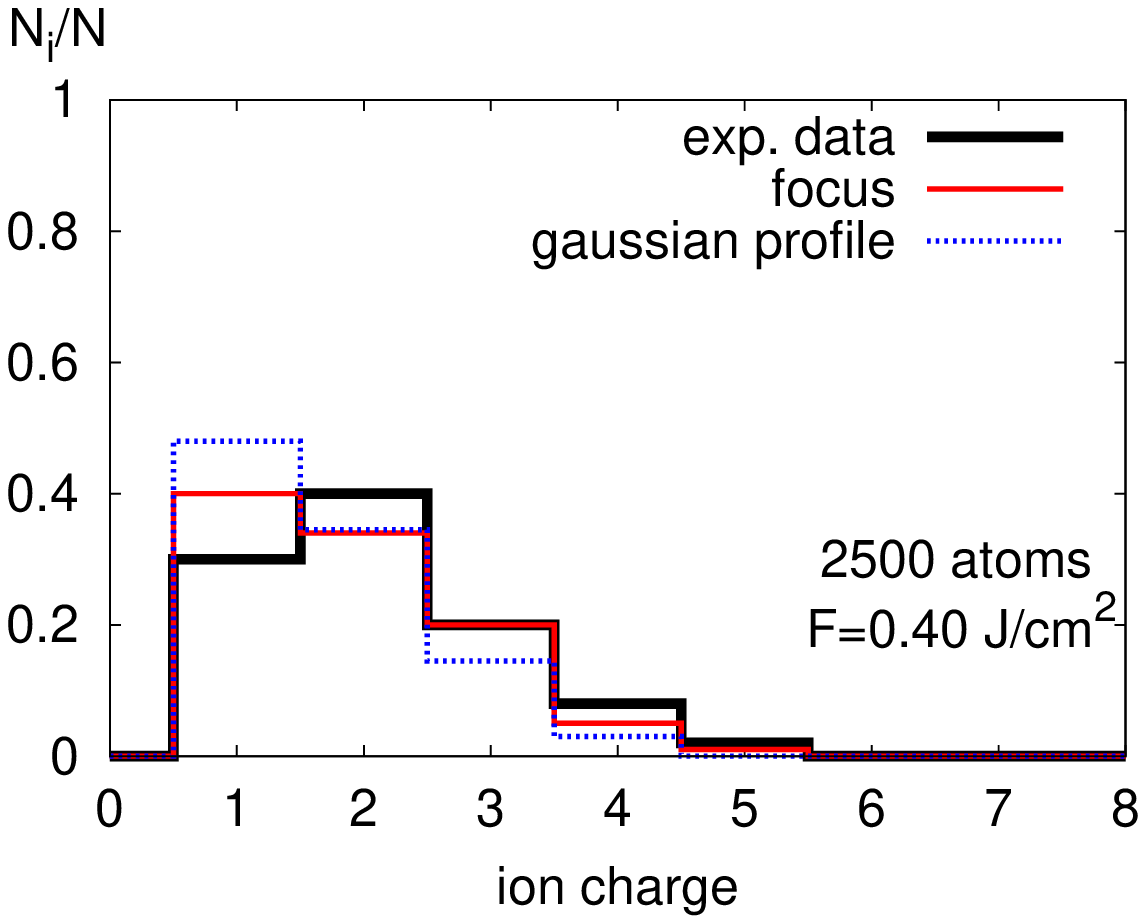}
d) \includegraphics[width=5cm]{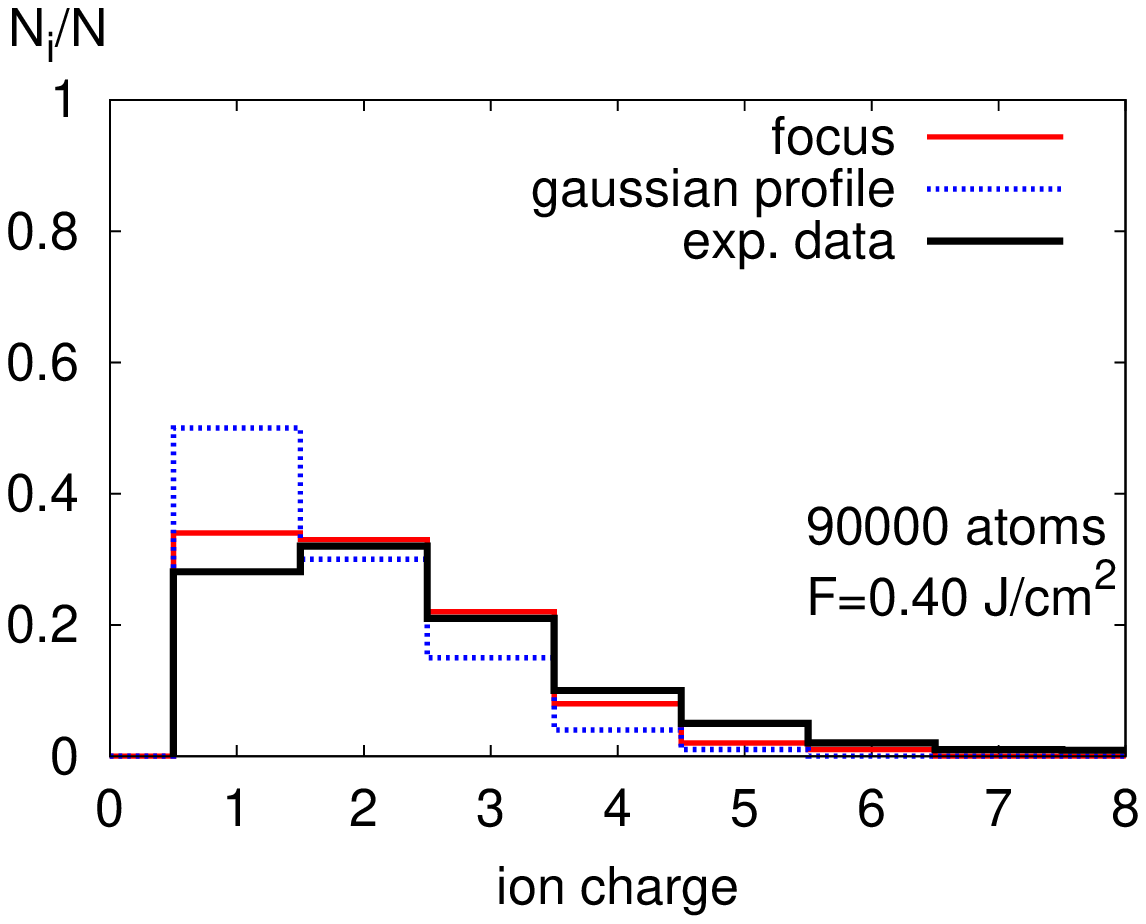}
f)\includegraphics[width=5cm]{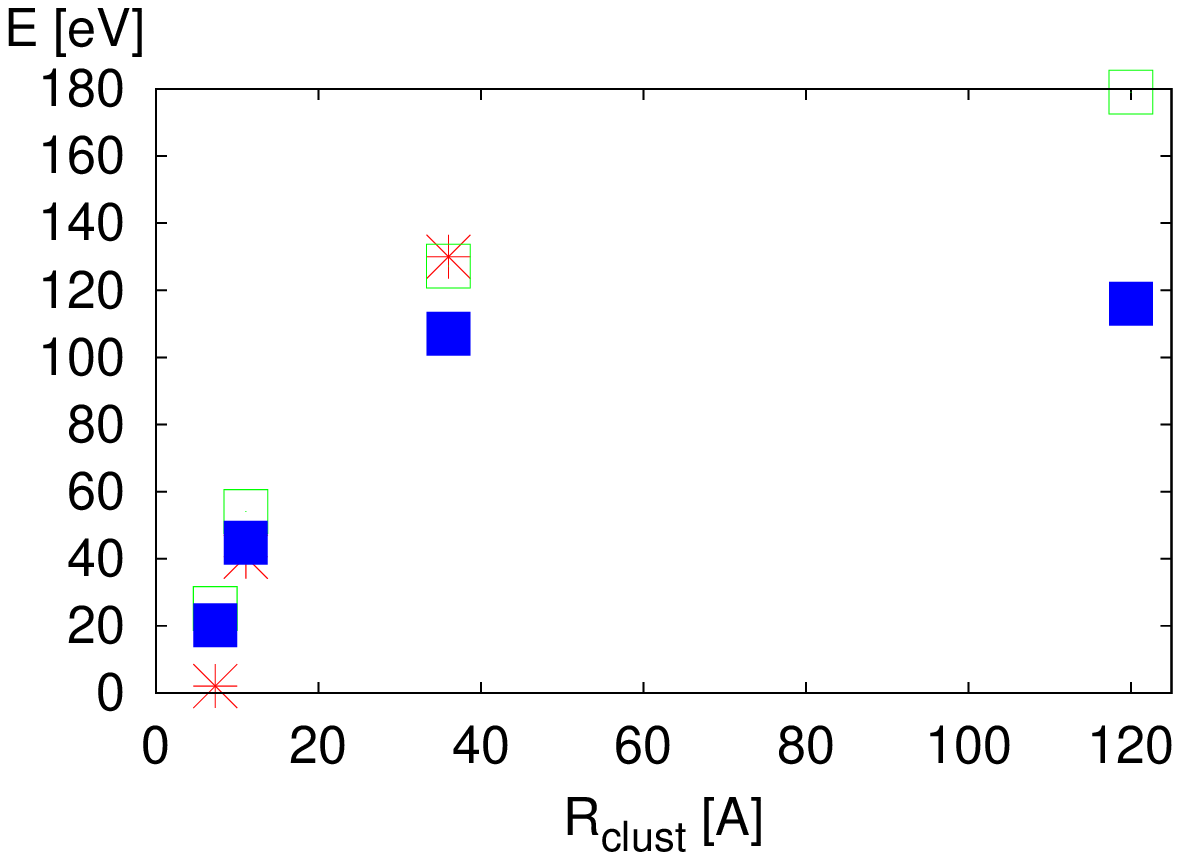}\\

\caption{\label{charge} Ion fractions, $N_i/N$, within the irradiated clusters estimated for various cluster sizes: (a) 20 atoms, (b) 70 atoms, (c) 2500 atoms and (d) 90000 atoms.The number, $N_i$ denotes here the number of $Xe^{+i}$ ions, and the number, $N$, is the total number of ions. Model predictions for clusters placed in the centre of the focussed beam and predictions integrated over the approximately estimated gaussian spatial pulse profile are shown. Also: (e) average charge, $Z$, created within the irradiated cluster, and (f) average energy absorbed per ion, $E$, within the irradiated cluster as a function of the cluster radius, $R$. The model estimates (squares) were obtained with pulses of different intensities and lengths but of fixed integrated radiation flux, $F=0.4$ J/cm$^2$, and then averaged over the number of pulses.  Experimental data are plotted with stars. Model predictions for clusters placed in the centre of the focussed beam (open squares) and predictions integrated over the approximately estimated spatial pulse profile (filled squares) are shown.
}
\end{figure*}

\begin{figure*}
a) \includegraphics[width=5cm]{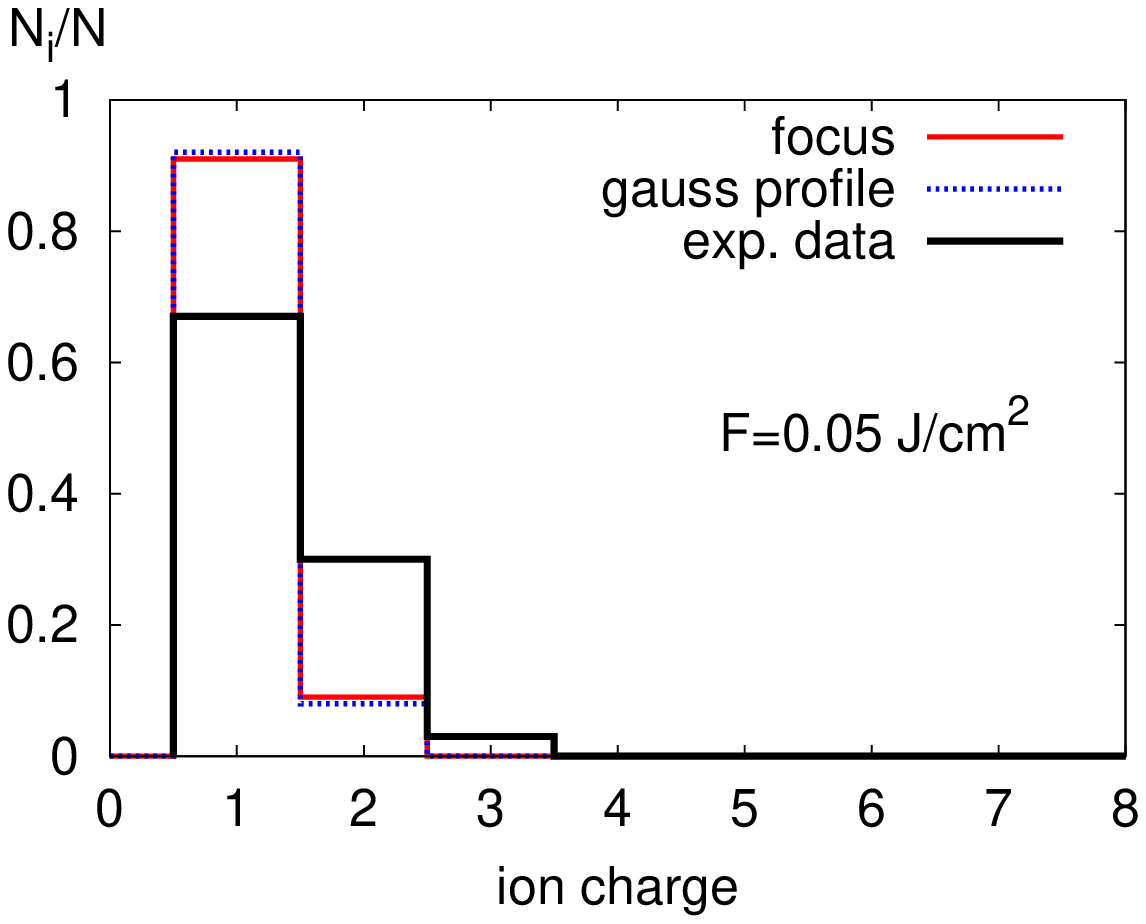}
c) \includegraphics[width=5cm]{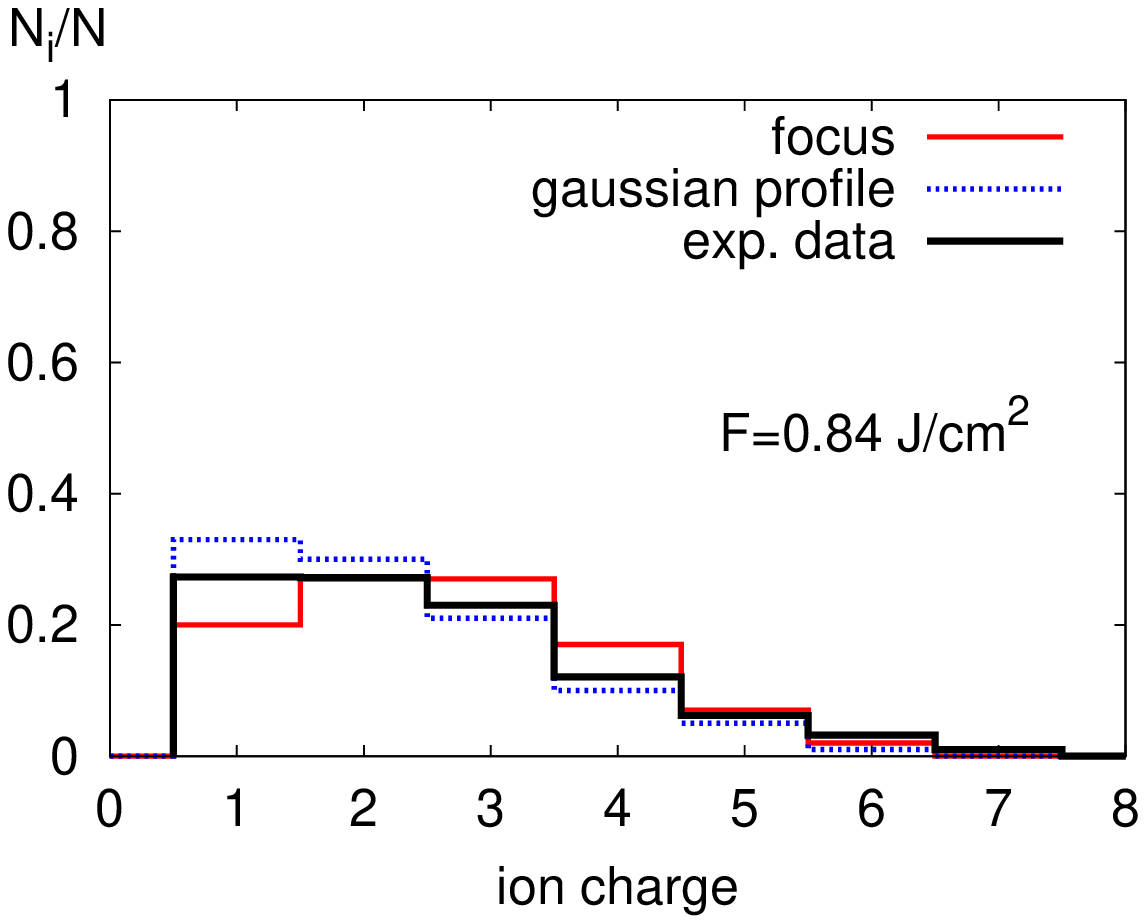}
e) \includegraphics[width=5cm]{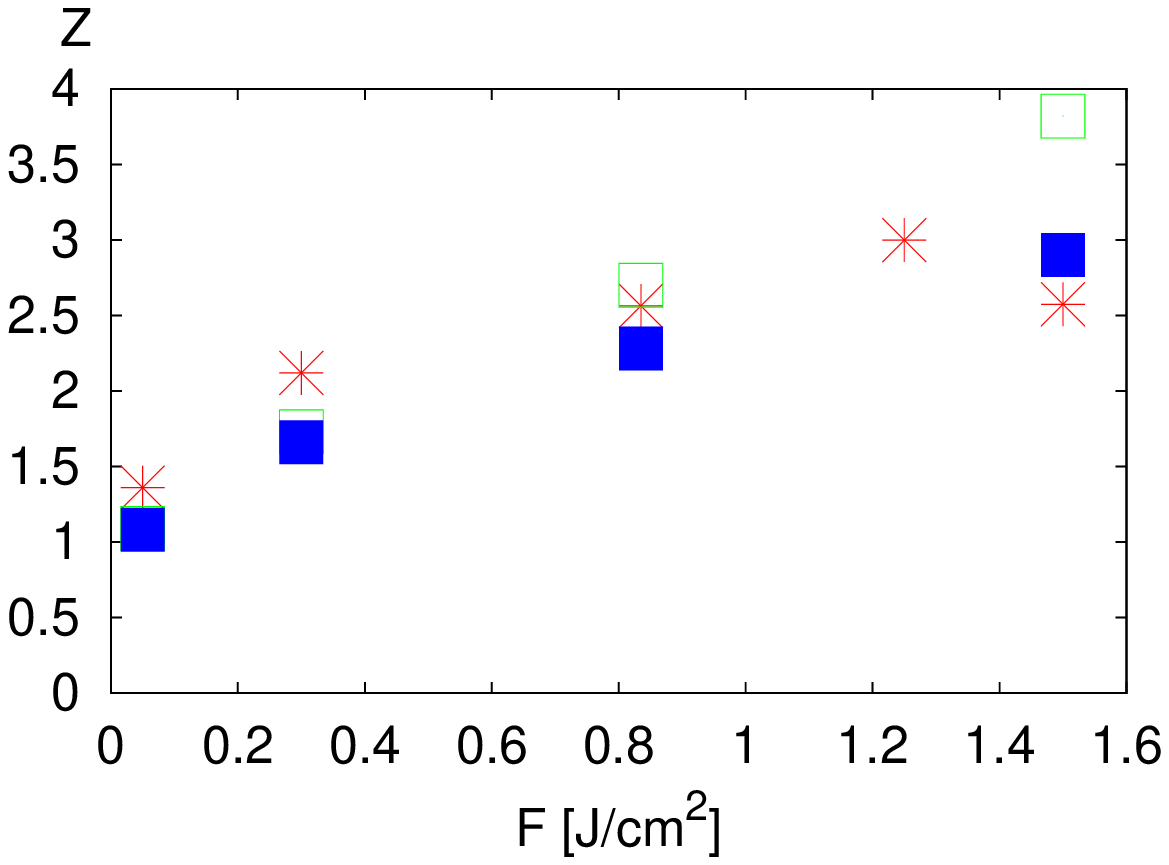}\\
b) \includegraphics[width=5cm]{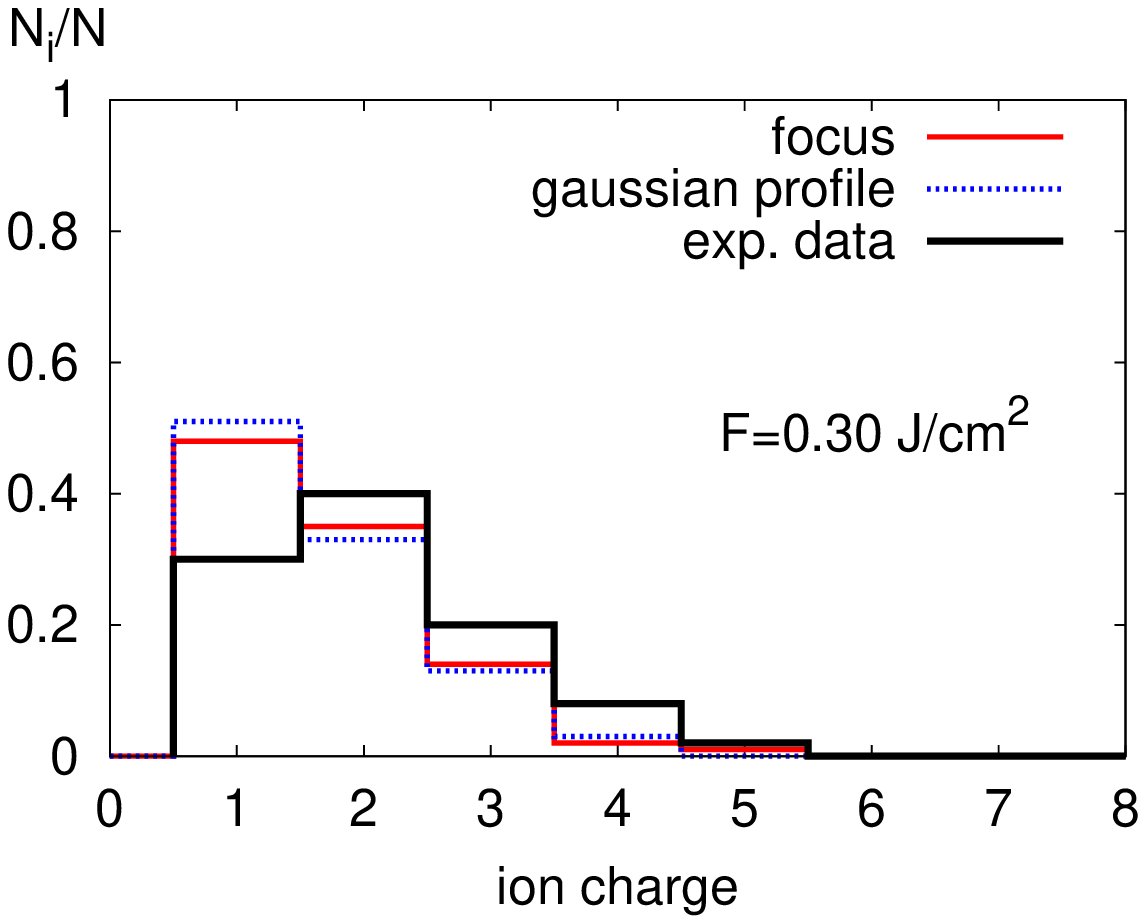}
d) \includegraphics[width=5cm]{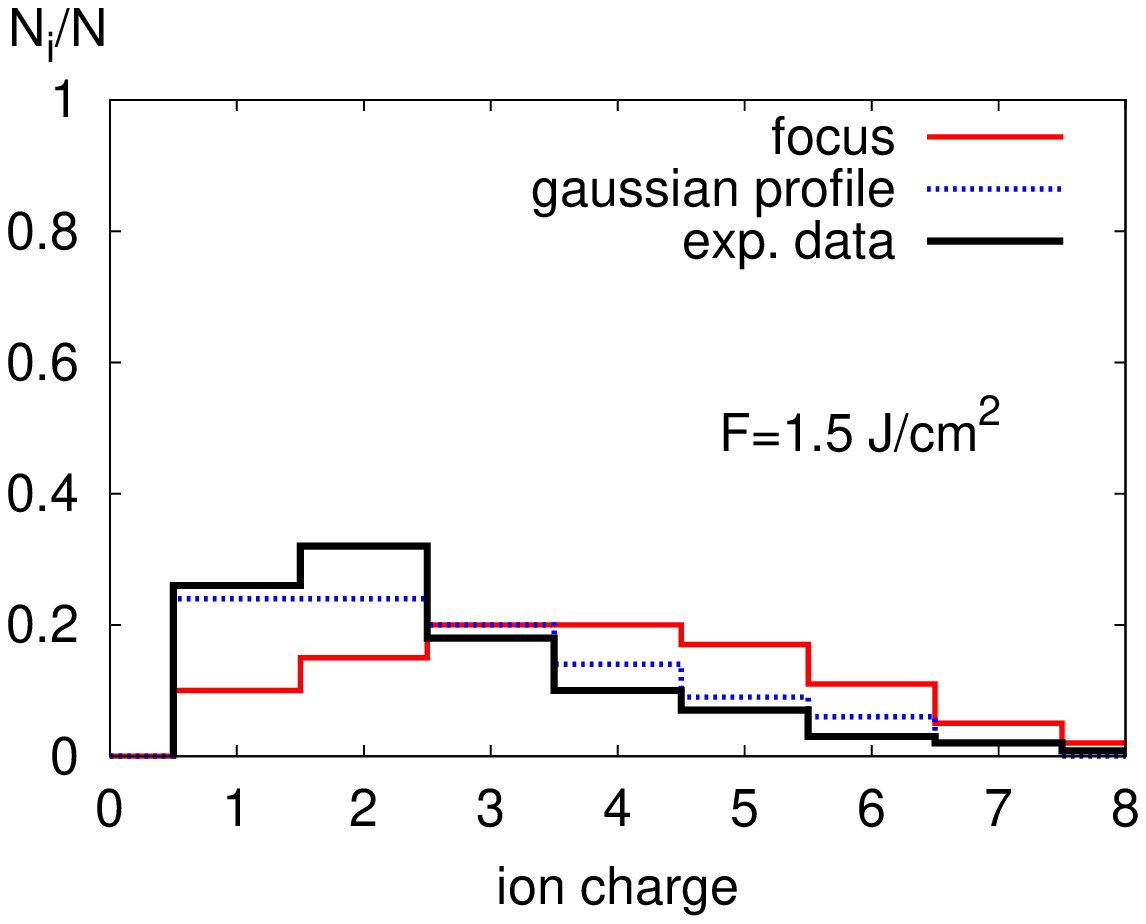}
f) \includegraphics[width=5cm]{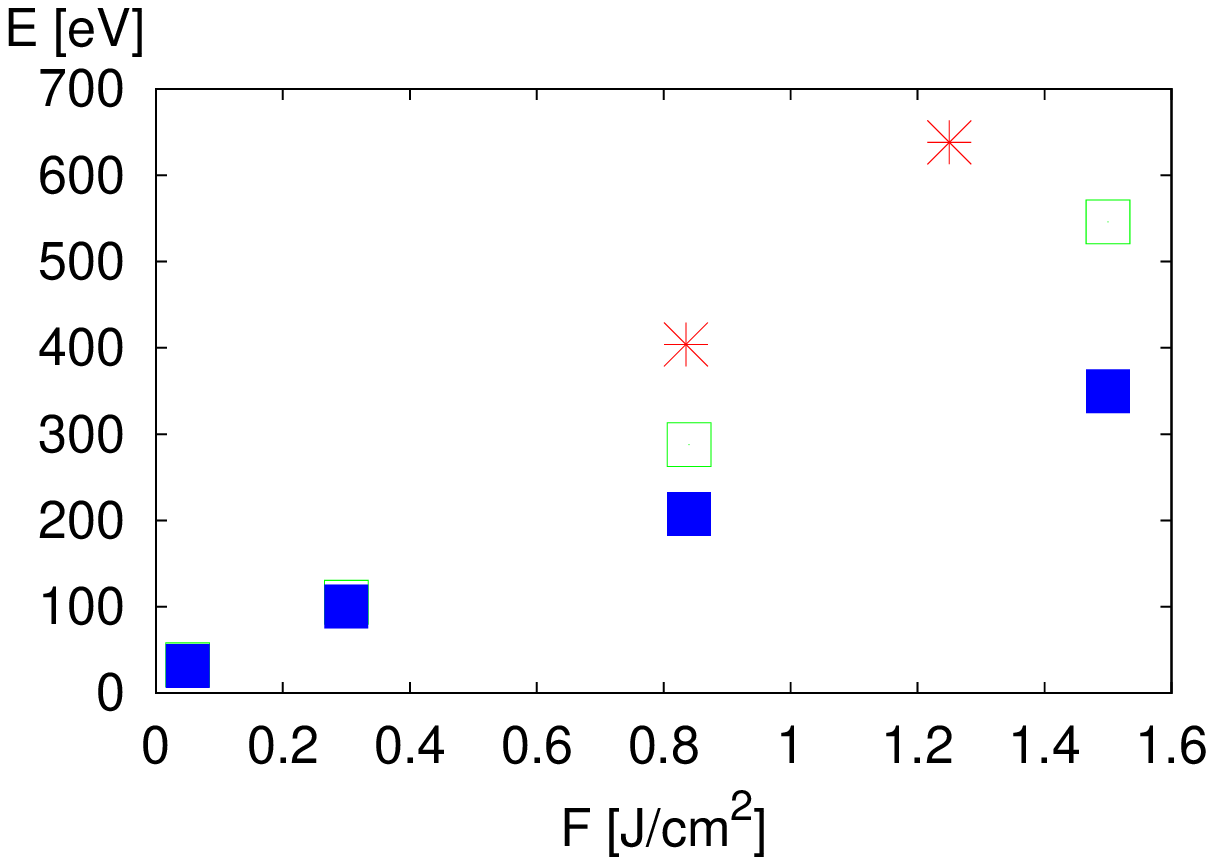}\\

\caption{\label{fcharge} Ion fractions, $N_i/N$, within the irradiated xenon clusters ($N_{atoms}=2500$). These clusters were irradiated with rectangular pulses  of fluences: (a) $F=0.05$ J/cm$^2$, (b) $F=0.3$ J/cm$^2$, (c) $F=0.84$ J/cm$^2$, and (d) $F=1.5$ J/cm$^2$. The number, $N_i$, denotes here the number of  $Xe^{+i}$ ions, and the number $N$ is the total number of ions. Model predictions for clusters placed in the centre of the focussed beam and predictions integrated over the approximately estimated gaussian spatial pulse profile are shown. Also: (e) average charge, $Z$, created within the irradiated xenon cluster ($N_{atoms}=2500$), and (f) average energy absorbed per ion, $E$, 
within the irradiated cluster as a function of the pulse fluence, $F$. The model estimates (squares) were obtained with pulses of different intensities and lengths but of fixed fluence, $F=0.05-1.5$ J/cm$^2$, and then averaged over the number of pulses. Experimental data are plotted with stars. Model predictions for cluster placed in the centre of the focussed beam (open squares) and predictions integrated over the approximately estimated spatial pulse profile (filled squares) are shown. 
}
\end{figure*}
\indent
Dynamics within irradiated clusters depends on the pulse fluence and cluster size. We performed simulations of irradiated clusters at the parameter values corresponding to those at the first FLASH cluster experiment at DESY at $100$ nm radiation wavelength \cite{desy,desy7}. These predictions will now be compared to the experimental data. 

The simulations were performed for clusters exposed to single rectangular VUV pulses of a fixed fluence, but of various intensities and pulse durations. The pulse intensity was $10^{12}-10^{14}$ W/cm$^2$ and the pulse length was, 
$\Delta t \leq 50$ fs.  
The predictions obtained from different events were then averaged over the number of events. This scheme followed the experimental data analysis: experimental data were obtained after averaging the single shot data obtained with FEL pulses of various temporal shapes but of a fixed radiation flux. 
In particular, the experimentally estimated histograms of ion charge and predictions on the average kinetic energy per ion were obtained from the averaged time-of-flight (TOF) spectra. The TOF detector could record charged particles (ions) only. The experimental ion intensities were obtained by integrating and averaging the TOF signal over subsequent FEL pulses. Some of those intensities were corrected for the relative geometric acceptances of the TOF detector or the MCP detector efficiencies for different charge states. The maximal experimental uncertainty of the fluence estimation is given by factor of $2.5$. We obtain predictions for two limiting cases: i) a cluster placed in the centre of the focussed beam and ii) the position of the cluster integrated over the approximately estimated spatial profile of the pulse. 
As we will see later, at higher pulse fluences averaging over the spatial pulse profile will have a higher impact on the model predictions, comparing to the
case of a cluster placed in the centre of focussed beam. This is due to the strong nonlinear dependence of the ion charge created within a cluster on the pulse fluence at higher pulse fluences. At lower fluences this dependence is linear.

First we will investigate ionization dynamics at different cluster sizes.
In figs.\ \ref{charge}a-d we plot the charge state distributions at the end of expansion phase obtained for four different cluster sizes irradiated at a fixed pulse fluence of $0.4$ J/cm$^2$.
The trend can be understood in the following way. The inverse bremsstrahlung is the dominant mechanism of energy absorption within our model. This process heats up the quasi-free electrons within the cluster. The hot electrons collide with ions and atoms, producing higher charges and releasing new electrons. These processes compete with the three-body recombination processes that reduce the ion charge and decrease the electron density. After the pulse is over, and the system reaches the local thermodynamic equilibrium (LTE) state, the number of ionization and recombination events saturates, changing slowly with the decreasing temperature within the cluster. 

Within small clusters a large fraction of electrons released during the ionization processes can leave the clusters early in the exposure. The remaining electrons are not heated efficiently via inverse bremsstrahlung processes due to their low density within the cluster. Consequently, only low charge states are observed (fig.~\ref{charge}a-b).  
In contrast, within large clusters of 2500 and 90000 xenon atoms only a small fraction of the released electrons are able to escape from the cluster. The width of the positively charged outer shell is small with respect to the radius of the neutral core. Electrons confined within the core are then heated efficiently. That leads to further collisional ionizations. 
In fig.~\ref{charge}c-d we plot the respective ion fractions. Our predictions are in agreement with experimental predictions. The trend of the size dependence is correct: at fixed pulse energy the maximal ion charge created increases with the cluster size until it saturates at larger cluster sizes. 
The average charge is plotted as a function of cluster radius in fig.\ \ref{charge}e. Similarly, at higher cluster radii the average charge saturates. This is in agreement with the experimental data. 
Saturation of the ion charge created within large clusters irradiated with a pulse of a fixed fluence is due to the fact that for clusters large enough the energy absorbed from the pulse (per atom) will not be sufficient for the creation of higher charge states.

Below we also show our estimates for the average kinetic energy per ion, $E$, as a function of the cluster radius (fig.~\ref{charge}f). The energy absorption per atom increases with the cluster size and also saturates for larger clusters. 

The dependence of the ionization dynamics on the radiation flux has been experimentally investigated for xenon clusters consisting of 2500 atoms.
Below we show the plots of the ion fractions obtained with the experimental data and the ion fractions obtained with our model (fig.\ \ref{fcharge}a-d).
Higher pulse fluences lead to the creation of higher charge states within the clusters in accordance with the mechanism described above.

At the lowest radiation flux, $F=0.05$ J/cm$^2$ (e. g. $I=5 \cdot 10^{12}$ W/cm$^2$ and $\Delta t=10$ fs for a rectangular pulse), charges up to +3 were found. Our model shows only up to doubly charged ions. As suggested in ref. \cite{ziajab2}, this small discrepancy between experiment and our predictions may be due to the fact that at low fluences the many-body recombination effects within the cold electron plasma can be important \cite{brabec}. They could then lead to higher ionization states that we do not observe within this model. 

At the flux, $F=0.3$ J/cm$^2$, experimental ion fractions have a maximum at Xe$^{+2}$, whereas our predictions peak at Xe$^{+1}$. However, the maximal ion charge is found to be +5 with both experimental data and simulation results.
At higher flux, $F=0.84$ J/cm$^2$ and the $F=1.5$ J/cm$^2$, the ion fractions are in a good agreement with experimental data. Maximal charges up to +7 and +8
are observed respectively.

The average charge is plotted as a function of radiation flux in fig.\ \ref{fcharge}e. With the increasing pulse fluence the average charge created increases. The charges calculated are very close to the corresponding experimental values. Also, as expected, the content of neutral atoms found 
within the cluster (not shown) decreases at the increasing pulse fluence.

Below we show also the average kinetic energy per ion (estimated with our model) as a function of the radiation flux (fig.\ \ref{fcharge}f). The sparse experimental data do not allow the identification of the trend of the fluence dependence, i.e. whether it is linear or non-linear. Our predictions slightly underestimate the experimental predictions at high fluences but stay within the error limit given by the experimental uncertainty of fluence estimation.

In summary, we used a microscopic model based on a first principle Boltzmann approach to investigate non-equilibrium dynamics within atomic clusters of various sizes irradiated by single VUV pulses.  This model is computationally efficient for small and large clusters, and includes various processes that are relevant at VUV photon energies. Predictions obtained at various cluster sizes and various pulse fluences were compared to the experimental data and found to be in good overall agreement, especially if one considers the experimental uncertainty of pulse fluence estimation given by a factor of $2.5$.
The results obtained have been cross-checked with independent molecular dynamics simulations, and have also showed good agreement.

The results obtained demonstrate the different ionization dynamics of small and large clusters. For small clusters the efficient escape of electrons from the cluster reduces the number of higher charges created within the sample during collisional ionizations. Recombination is also suppressed due to the absence of free electrons. For large clusters, many electrons stay within the sample as they are kept there by the attractive potential of the outer shell. They are heated by the IB process, and ionize and recombine efficiently. Higher charges are created, and as a result of efficient recombination neutral atoms appear within the cluster core. After the pulse is over, the system reaches the LTE, and the total number of ions and atoms within the sample changes only weakly.
A fraction of neutral atoms remains within the core, its magnitude depending on the cluster size. This indicates that the ions detected during the experiment come mainly from the surface and the outer part of the cluster.  


In conclusion, we find that formation of high charge states and the strong absorption of VUV radiation is quantitatively understood within the framework of our model.

\begin{acknowledgments}
Beata Ziaja is grateful to Christoph Bostedt, Henry Chapman, Cornelia Deiss, Tim Laarmann, Adrian Mancuso, Wojciech Rozmus, Robin Santra and Abraham Sz\"oke for illuminating comments. Authors thank the colleagues from Dresden, especially Jan Michael Rost and Ulf Saalmann, as well as Joshua Jortner (Tel Aviv) for fruitful discussions. 
This research was supported by the German Bundesministerium f\"ur Bildung und Forschung with grants No.\ 05 KS4 KTC/1, No.\ 05 KS7 KT1 and by the Helmholtz
Gemeinschaft, Impulsfond VH-VI-302. 
\end{acknowledgments}


\begin{thebibliography}{10}

\bibitem{desy2006}
DESY.
\newblock {\em {Technical Design Report of the European XFEL, DESY, ISBN
  3-935702-17-5}}, 5:7--9, 2006.

\bibitem{slac}
LCLS.
\newblock {\em {LCLS: The First Experiments., SSRL, SLAC, Stanford, USA}},
  2000.

\bibitem{jap}
T.~Shintake and SCSS Team.
\newblock {Status of Japanese XFEL Project and SCSS test accelerator}.
\newblock {\em {Proceedings of FEL 2006, BESSY, Berlin, Germany}}, pages
  33--36, 2006.

\bibitem{xfelinfo2007}
DESY.
\newblock {XFEL}-info.
\newblock {\em DESY, {http://xfelinfo.desy.de}}, 5:18--25, 2007.

\bibitem{l1}
R.~Neutze, R.~Wouts, D.~van~der Spoel, E.~Weckert, and J.~Hajdu.
\newblock {\em Nature}, 406:752--757, 2000.

\bibitem{miao}
J.~Miao, K.~Hodgson, and D.~Sayre.
\newblock {\em Proc. Natl. Acad. Sci.}, 98:6641, 2001.

\bibitem{gyula1}
Z.~Jurek, G.~Oszl\'anyi, and G.~Faigel.
\newblock {\em Europhys. Lett.}, 65:491, 2004.

\bibitem{plasma4}
{S. P. Hau-Riege et al.}
\newblock {\em Phys. Rev. E}, 71:061919, 2005.

\bibitem{liver1en}
{S. P. Hau-Riege et al.}
\newblock {\em Phys. Rev. Lett.}, 98:198302, 2007.

\bibitem{chapman}
{H. Chapman et al.}
\newblock {\em Nature Physics}, 2:839, 2006.

\bibitem{desy}
{H. Wabnitz et al.}
\newblock {\em Nature}, 420:482, 2002.

\bibitem{desy5}
{J. Schulz et al.}
\newblock {\em Nucl. Instr. and Meth. in Phys. Res. B}, 507:572, 2003.

\bibitem{desy2}
H.~Wabnitz.
\newblock {Interaction of intense VUV radiation from FEL with rare gas atoms
  and clusters}.
\newblock {\em Doctoral Thesis}, DESY-THESIS-2003-026, 2003.

\bibitem{desy7}
{T. Laarmann et al.}
\newblock {\em Phys. Rev. Lett.}, 92:143401, 2004.

\bibitem{santra}
R.~Santra and C.~H. Greene.
\newblock {\em Phys. Rev. Lett.}, 91:233401, 2003.

\bibitem{santra1}
Z.~B. Walters, R.~Santra, and C.~H. Greene.
\newblock {\em Phys. Rev. A}, 74:043204, 2006.

\bibitem{siedschlag}
C.~Siedschlag and J.M. Rost.
\newblock {\em Phys. Rev. Lett.}, 93:043402, 2004.

\bibitem{georg}
I.~Georgescu, U.~Saalmann, and J.-M. Rost.
\newblock {\em Phys. Rev. A}, 76:043203, 2007.

\bibitem{bauer}
D.~Bauer.
\newblock {\em J. Phys. B}, 37:3085, 2004.

\bibitem{brabec}
C.~Jungreuthmayer, L.~Ramunno, J.~Zanghellini, and T.~Brabec.
\newblock {\em J. Phys. B}, 38:3029, 2005.

\bibitem{rusek}
M.~Rusek and A.~Orlowski.
\newblock {\em Phys. Rev. A}, 71:043202, 2005.

\bibitem{ziajab2}
B.~Ziaja, H.~Wabnitz, E.~Weckert, and T.~M\"oller.
\newblock {\em New J. Phys}, 10:043003, 2008.

\bibitem{ziajab3}
B.~Ziaja, H.~Wabnitz, E.~Weckert, and T.~M\"oller.
\newblock {\em Europhys. Lett.}, 82:24002, 2008.

\bibitem{desy8}
{M. Hoener et al.}
\newblock {\em J. Phys. B: At. Mol. Opt. Phys.}, 41:181001, 2008.

\end{thebibliography}

\end{document}